\documentclass[prb,aps,twocolumn,superscriptaddress,floatfix,showpacs]{revtex4}
\usepackage{graphicx}
\usepackage{amssymb}
\usepackage{amsmath}
\usepackage{xspace}
\usepackage{color}
\usepackage{soul}

\begin{document}

\title{Spin dynamics in paramagnetic diluted magnetic semiconductors}
\author{Van-Nham Phan}
\affiliation{Institute of Research and Development, Duy Tan University, K7/25 Quang Trung, Danang, Vietnam}
\author{Minh-Tien Tran}
\affiliation{Institute of Research and Development, Duy Tan University, K7/25 Quang Trung, Danang, Vietnam}
\affiliation{Institute of Physics, Vietnam Academy of Science and Technology, 10 Daotan, Hanoi, Vietnam}

\begin{abstract}
Microscopic properties of low-energy spin dynamics in diluted magnetic semiconductor are addressed in a framework of the Kondo lattice model including random distribution of magnetic dopants. Based on the fluctuation-dissipation theorem, we derive an explicit dependence of the spin diffusion coefficient on the single-particle Green function which is directly evaluated by dynamical mean-field theory. In the paramagnetic state, the magnetic scattering has been manifested to suppress spin diffusion. In agreement with other ferromagnet systems, we also point out that the spin diffusion in diluted magnetic semiconductors at small carrier concentration displays a monotonic $1/T$-like temperature dependence. By investigating the spin diffusion coefficient on a wide range of the model parameters, the obtained results have provided a significant scenario to understand the spin dynamics in the paramagnetic diluted magnetic semiconductors.
\end{abstract}

\maketitle
\date{\today}
\section{Introduction}

Recently, diluted magnetic semiconductors (DMSs) have attracted much attention due to their prospective applications in spin electronics (spintronics)~\cite{Wolf}, where the integration of data processing and the magnetic storage are incorporated into a single chip.~\cite{RMP141,RMP142} In typical DMS materials, magnetic ions are doped into a semiconducting host, for example, manganese ions Mn$^{2+}$ are doped into semiconducting host GaAs.~\cite{Ohnos}
They play a dual role of both an acceptor and localized magnetic moment due to their partially filled $d$ shell.~\cite{Fed} At low temperature and with small doping, the DMSs have often been found to be ferromagnetic, for instance, Ga$_{1-x}$Mn$_x$As is ferromagnetic at $x\sim $1$\div$ 7$\%$.~\cite{Dietl,Chatt} In this case, the spin-spin correlation is long range, and the spin excitation is mediated by the spin wave states. In the opposite case, i.e., in the paramagnetic state, the spin-spin correlation is short range and the spin polarization of carriers in general is not spatially uniform, a hydrodynamic spin diffusion thus is induced in the system. To implement DMSs for real spintronics impact, the understanding of the spin dynamics is important.
The magnetic relaxation time $\tau(k)$ under a magnetic disturbance with wave vector $k$ can be determined through the spin diffusion constant $D_s$ by $\tau(k)=1/D_sk^2$.~\cite{Fish02} Therefore the spin diffusion constant is the principal quantity that provides information on the spin dynamics. When the system reaches the paramagnetic-ferromagnetic phase transition point, the spin diffusion constant decreases and the process of spin relaxation slows down. The magnetic relaxation time and the spin fluctuations therefore are dominant at small wave vector of the magnet excitation spectrum in the paramagnetic state. In degenerate systems, the spin diffusion constant can be determined from the spin conductivity at zero frequency by the Einstein relation. The spin diffusion constant in the paramagnetic state at high temperature was first studied by de Gennes.~\cite{De} Following his work, various models have been applied to study the spin diffusion constant in magnetic systems such as the $t-J$ model,~\cite{Bon} the Heisenberg model,~\cite{Huber} or the double exchange model.~\cite{Cher}

In this paper, we also present a microscopic calculation of the spin diffusion constant of DMSs.
In DMSs, the doped magnetic ions act as an acceptor, so the main charge carrier in the DMSs is the hole. The holes are assumed to be able to hop in the lattice that creates the quasiparticle band and, for simplicity, it can be modeled by the tight-binding approximation.~\cite{BB01,RMP06}
The local spin exchange between the magnetic ions and the holes is the essential ingredient of the magnetic properties of the DMSs.~\cite{Chatt,HS05}
We assume that the doped magnetic ions are randomly substituted in the cation sites, so in the lattice only a fraction of lattice sites is occupied by the magnetic impurities and the remaining sites are nonmagnetic. The local spin exchange is valid only on the magnetic impurity sites.
Due to the random chemical substitution, a random local potential for the charge carriers
at the magnetic impurity sites  is also taken into account. This situation looks similar to doped manganites in which rare-earth ions are replaced by divalent alkaline ions.~\cite{Tr03,LF01}
In doped manganites, a Falicov-Kimball term has succeeded in modeling the randomness of the doped ions.~\cite{Tr03,LF01} In the framework of dynamical mean-field theory (DMFT), a moderate disorder solution of the dc electrical resistivity has shown agreement with the experimental data observed for manganites, particularly in the paramagnetic state.~\cite{LF01} The DMFT has extensively been used for investigating strongly correlated electron systems.\cite{George} It is based on the fact that the self-energy depends only on frequency in the infinite dimensional limit. By adapting also the DMFT, the spin diffusion at any temperature down to the ferromagnetic transition point in manganites has been calculated.~\cite{Cher} A qualitative agreement of the theoretically calculated spin diffusion constant with the experimental data was also observed.~\cite{Cher} Developing from these achievements, in the present work we focus on the spin dynamical properties in the DMSs based on the DMFT. We construct a microscopic model for DMSs, in which the randomness of the doped magnetic ions is taken into account. The spin diffusion constant is related to the single-particle Green function and it can be evaluated within the framework of the DMFT. It is found that the spin diffusion enhances if the Fermi level settles inside the magnetic impurity band. Both magnetic scattering and temperature suppress the spin diffusion in the paramagnetic state. At small carrier concentrations, the spin diffusion displays a monotonic $1/T$-like temperature dependence. In contrast, for large carrier concentrations we find a minimum point at an intermediate temperature that is attributed to an occurrence of the low-energy short lived many body states in the system.

The present paper is organized as follows. In Sec.~II, we present a microscopic Hamiltonian essentially applied for the DMS materials and its DMFT solution in the paramagnetic state. Section~III outlines some general steps to derive the Einstein relation between the spin diffusion constant and spin conductivity through the fluctuation-dissipation theorem. In Sec.~IV, we present the numerical results and their discussions. A summary and conclusion are presented in the last section.

\section{Microscopic model for diluted magnetic semiconductors and its dynamical mean-field theory}

In the presence of random magnetic ions and their spin exchange with the itinerant carriers in DMSs, we construct the following Hamiltonian in the tight-binding approximation
\begin{align}
\mathcal{H}=&-t\sum_{\langle i,j\rangle\sigma}c_{i\sigma}^{\dagger}c^{}_{j\sigma}+2J\sum_{i}\alpha_{i}{\bf S}_{i}{\bf s}_{i}\nonumber\\
&-\mu\sum_in_i-\sum_{i}U\alpha_in_i,
\label{Hami}
\end{align}
where $c^{\dagger}_{i\sigma}$ ($c^{\null}_{i\sigma}$) is the creation (annihilation) operator for an itinerant carrier with spin $\sigma$ at lattice site $i$. The first term in the Hamiltonian~(\ref{Hami}) represents the tight-binding model for the itinerant carriers in DMSs. $t$ is the hoping integral, and in the limit $d\rightarrow \infty$, it scales with the spatial dimension $d$ as $t=t^\ast/2\sqrt d$.~\cite{George} In the following, we will take $t^\ast=1$ as the unit of energy.~\cite{George} ${\bf S}_{i}$ is the spin of the magnetic impurity at lattice site $i$, while ${\bf
s}_{i}=\sum_{ss'}c^{\dagger}_{is}\boldsymbol{\sigma}_{ss'}c^{\null}_{is'}/2$ is the spin of the itinerant carriers ($\boldsymbol{\sigma}$ are the Pauli matrices). $J$ is the strength of the local spin exchange.
$U$ is a local disorder strength, which reflects the energy difference, when the lattice site is occupied by magnetic ions. The chemical potential $\mu$ is  introduced to control carrier doping.
In the Hamiltonian~(\ref{Hami}), we have also included $\alpha_i$ as a classical variable that takes the value of either $1$ or $0$ if site $i$ is occupied or unoccupied respectively by magnetic ion. The introduced variable $\alpha_i$ ensures that the spin exchange and the local disorder are valid only on the lattice sites that are occupied by the magnetic ions. Its distribution function is binary:
\begin{equation}
P(\alpha)=(1-x)\delta(\alpha)+x\delta(1-\alpha),
\end{equation}
where $x$ is the doping number of the magnetic ions. In the case of $\alpha_i=1$ for all $i$, the first three terms in the Hamiltonian illustrate the Kondo lattice model.~\cite{Chatt} Because of the large localized spin of magnetic ions, for instance, $S=5/2$ for an Mn ion at half-filling, the localized magnetic spin can be considered to behave classically, as widely assumed.~\cite{Chatt,MAD02}

The Hamiltonian~\eqref{Hami} looks analogous to the double exchange model with diagonal disorder. The later has been solved successfully by the DMFT to reveal a complex phase structure in doped manganites.~\cite{Tr03,PT05} Without the random disorder, the DMFT has also been adapted to calculate the critical temperature for the ferromagnetic transition in DMSs.~\cite{Chatt} Meanwhile, in the presence of the disorder potential due to the doping of magnetic ions, the
transport properties in DMSs have been studied by the DMFT as well.~\cite{HS05} In the present work, we make further use of the DMFT to investigate the spin dynamics in DMSs.

The key point of the DMFT lies in the limit of infinite space dimensions. In this limit, the self-energy is local and does not depend on momentum. The local Green function of itinerant carriers can be determined via the Dyson equation
\begin{equation}
\tilde{G}(i\omega_n)=\int d\varepsilon \rho(\varepsilon)\frac{1}{i\omega_n -\varepsilon +\mu -
\tilde{\Sigma}(i\omega_n)},
\label{Green1}
\end{equation}
where $\omega_n=(2n+1)\pi T$ is the Matsubara frequency at temperature $T$, $\tilde{\Sigma}(i\omega_n)$ is the self-energy, and
$\rho(\varepsilon)$ is the noninteracting density of states (DOS) of the itinerant carriers. Without loss of generality, we use the semicircular DOS defined by $\rho(\varepsilon)=\sqrt{(4-\varepsilon^2)}/2\pi$. The Green function and self-energy in Eq.~\eqref{Green1} have been written in the spin matrix form denoted by tildes.

The self-consistency of the DMFT requires that the local Green function in Eq.~(\ref{Green1}) must coincide with the Green function determined within the dynamics of the effective single impurity embedded in the dynamical mean-field medium
\begin{equation}\label{G001}
\tilde G(i\omega_n)=\beta\frac{\partial \mathcal{F}}{\partial \tilde{\mathcal{G}}^{-1}(i\omega_n)},
\end{equation}
where $\mathcal{F}=-T\int d\alpha P(\alpha)\ln \mathcal{Z}_{\textrm{eff}}(\alpha)$ is the free energy of the system, $\tilde{\mathcal{G}}(i\omega_n)$ is a Green function representing the dynamical mean field, and $\beta=1/T$. $\mathcal{Z}_{\textrm{eff}}(\alpha)$ is the partition function of the effective single impurity:
\begin{equation}\label{Zeff}
\mathcal{Z}_{\rm{eff}}(\alpha)=\int d\Omega_{\bf{m}}e^{-S_{\textrm{eff}}(\bf{m},\alpha)},
\end{equation}
where
\begin{equation}
S_{\rm{eff}}({\bf m},\alpha)=-\sum_n\ln \det[\tilde {\cal G}^{-1}(i\omega_n)-J{\bf m} \cdot \boldsymbol{\sigma}\alpha +U\alpha]
\end{equation}
is the action. The integral~\eqref{Zeff} is taken over all possible values, ${\bf m}=(m_x,m_y,m_z)$, of the classical localized magnetic moment $\mathbf{S}$. Changing the integral variable in Eq. (\ref{Zeff}) into spherical coordinates, we obtain an explicit expression for the local Green function defined in Eq.~\eqref{G001}:
\begin{align}
G_{\sigma}(i\omega_n)=&2\pi\int_{-1}^{1}dy\int d\alpha \frac{1}{Z_{\textrm{eff}}(\alpha)}P(\alpha)\nonumber\\
&\times\exp\sum_n \ln[\Gamma_\uparrow^\alpha(i\omega_n)\Gamma_\downarrow^\alpha(i\omega_n) -J^2(1-y^2)]\nonumber\\
&\times\frac{\Gamma_{-\sigma}^\alpha(i\omega_n)}{\Gamma_\uparrow^\alpha(i\omega_n)\Gamma_\downarrow^\alpha(i\omega_n)-J^2(1-y^2)},
\label{Green2}
\end{align}
where 
\begin{equation}
\Gamma_\sigma^\alpha(i\omega_n)=\mathcal{G}^{-1}_{\sigma}(i\omega_n)+J\sigma y\alpha+U\alpha.
\end{equation}
The self-energy in Eq.~\eqref{Green1} can be determined from the Dyson equation
\begin{eqnarray}
\tilde{\Sigma}(i\omega_n)=\tilde{\mathcal{G}}^{-1}(i\omega_n)-\tilde{G}^{-1}(i\omega_n).
\label{Green3}
\end{eqnarray}
From Eqs.~(\ref{Green1}), (\ref{Green2}), and (\ref{Green3}), we obtain a set of self-consistent equations, which determine the self-energy and the lattice Green function.

\section{Spin-diffusion coefficient}

We start by noting that, in the paramagnetic state, the hydrodynamic description can be applied to address spin relaxation in our model.~\cite{Forster} In the present work, the spin-diffusion coefficient is calculated by employing the exact spectral representation for the spin-spin correlation function.
In the paramagnetic state, spins of carriers might arrange in a slightly inhomogeneous way that leads to a small gradient of the magnetization, for instance, in the $z$ direction. A slow spin current depending on time thereby exists in the sample. Following Fick's law, the relation between the spin current ${\bf j}_s$ and the gradient of  magnetization $m$ reads~\cite{Bennett}
\begin{equation}
{\bf j}_s=-D_s\nabla m =-D_s\chi \nabla B,
\end{equation}
where $D_s$ is the spin diffusion coefficient, $\chi$ is the static magnetic susceptibility, and $B$ is the $z$ component of the magnetic field. The spin diffusion process is associated with low frequency and long-wavelength excitations. In this regime, the spin-spin correlation function in (${\bf q},\omega$) space obtains quasielastic hydrodynamic diffusion behavior. According to the fluctuation-dissipation theorem, the spin-spin correlation function $S({\bf q},\omega)$ can be written in the hydrodynamic diffusion regime as following~\cite{Forster}:
\begin{equation}\label{7}
S({\bf q},\omega)\approx \frac{2}{1-e^{-\beta \omega}}\frac{\omega D_sq^2\chi({\bf q})}{\omega^2+(D_sq^2)^2}.
\end{equation}
Here, we have assumed cubic symmetry and spin-rotational invariance for the proposed model. $\chi({\bf q})$ is the momentum dependence of the static spin susceptibility. On the other hand, using the spectral representation, the spin-spin correlation function relates to the spin susceptibility by
\begin{equation}\label{8}
S({\bf q},\omega)= \frac{2}{1-e^{-\beta \omega}}\chi({\bf q},\omega),
\end{equation}
where $\chi({\bf q},\omega)$ is the dynamical spin susceptibility. Combining the two expressions of $S({\bf q},\omega)$ in Eqs.~\eqref{7} and \eqref{8}, we obtain
\begin{equation}\label{chiq0}
\chi ({\bf q},\omega)=\frac{\omega D_s{\bf q}^2\chi({\bf q})}
{\omega^2+(D_s{\bf q}^2)^2}.
\end{equation}
In the following, we will establish a relation between the dynamical spin susceptibility $\chi ({\bf q},\omega)$ and the spin current-current correlation function, where the latter can be evaluated within the Greenwood formalism.~\cite{Mahan} From the definition of the dynamical spin susceptibility
\begin{align}
\chi({\bf q},\omega)&=\frac{1}{2}\sum_{ij}\int dte^{i\omega t-i{\bf q}({\bf R}_i-{\bf R}_j)}
\langle [S^\alpha_i(t),S^\alpha_j(0)]\rangle\nonumber\\
&=\frac{1}{2}\langle [S^\alpha({\bf q},\omega),S^\alpha(-{\bf q},-\omega)]\rangle,
\end{align}
we can express it via the spin current correlation function
\begin{eqnarray}\label{chiq1}
\chi({\bf q},\omega)=\frac{1}{2}\frac{q^2}{\omega^2}\langle [j^\alpha({\bf q},\omega),
j^\alpha(-{\bf q},-\omega)]\rangle,
\end{eqnarray}
where $j^\alpha({\bf q},\omega)$ is the $\alpha$ component of the spin current, written in $({\bf q},\omega)$ space. Here we have used the continuous equation~\cite{Bon}
\begin{equation}
-\omega S^\alpha({\bf q},\omega)+q j^\alpha({\bf q},\omega)=0.
\end{equation}
The microscopic spin current is defined by the commutator of the Hamiltonian with the total spin polarization ${\bf P}^\alpha=\sum_i{\bf R}_i({S}^\alpha_i+{s}^\alpha_i)$.~\cite{Cher} One can notice that the Hamiltonian~\eqref{Hami} does not contain any direct coupling between the localized moments, hence
the spin current actually is the spin current of the itinerant carriers only, i.e., $j^{\alpha}({\bf q},t)=\sum_{{\bf k},ss'}v({\bf k})c^{\dagger}_{{\bf k}s}(t) \sigma^{\alpha}_{ss'}c_{{\bf k-q},s'}(t)$.
In a similar way to the calculation of  the particle or heat current-current correlation function,~\cite{Nham} within the Greenwood formalism,
the dynamical spin conductivity can be expressed in term of the spin current-current correlation function~\cite{Mahan}
\begin{equation}\label{sig0}
\sigma^\alpha_s({\bf q},\omega)= \textrm{Im}\frac{\Pi^\alpha({\bf q},\omega)}{\omega},
\end{equation}
where $\sigma^\alpha_s({\bf q},\omega)$ is the dynamical spin conductivity, and
$\Pi^\alpha({\bf q},\omega)=\frac{1}{2}\int dte^{i\omega t} \langle [j^\alpha({\bf q},t),j^\alpha(-{\bf q},0)]\rangle$ is the spin current-current correlation function. From the time-dependent spin current operator, in the zero-frequency and long-wavelength limit, we obtain the static spin conductivity
\begin{align}
\sigma_s&=\sigma^\alpha_s({\bf q}\rightarrow 0,\omega\rightarrow 0)\nonumber\\
&=\pi\sum_\sigma\int d\epsilon v^2(\varepsilon)\rho(\varepsilon)
\int d\omega' A^2_{\sigma}(\varepsilon,\omega')\left(-\frac{\partial f(\omega')}{\partial \omega'}\right),
\label{spinc}
\end{align}
which has been written in the unit of conductivity defined in Ref.~\onlinecite{ThPruschke}. Here $f(\omega)=1/[\exp(\omega/T)+1]$ is the Fermi distribution function. $A_{\sigma}(\varepsilon,\omega)$ is the spectral function of the itinerant carriers, i.e., $A_{\sigma}(\varepsilon,\omega)=\textrm{Im}G_\sigma(\varepsilon,\omega-i0^+)/\pi$. In the Bethe lattice, the current vertex $v(\varepsilon)$ in Eq.~\eqref{spinc} is $v(\varepsilon)=\sqrt{4-\varepsilon^2}$.~\cite{Freericks1998,Chat00} Note that from Eq.~\eqref{chiq1} we also have
\begin{equation}\label{19}
\chi({\bf q},\omega)= \frac{1}{2}\frac{q^2}{\omega^2}\Pi^\alpha({\bf q},\omega).
\end{equation}
Combining Eqs.~\eqref{19} and~\eqref{chiq0} with the help of Eq.~\eqref{sig0} in the limit $({\bf q}\rightarrow 0,\omega\rightarrow 0)$, we arrive at a formal expression of the Einstein relation which relates the spin conductivity and the spin-diffusion coefficient
\begin{equation}\label{Ds}
D_s\chi=\sigma_s.
\end{equation}
In this way, the spin diffusion coefficient has been expressed in term of the spin conductivity. In a next section, we will discuss the spin dynamics scenario through the spin diffusion coefficient instead of the spin conductivity.

\section{Numerical results}

From the Einstein relation in Eq.~\eqref{Ds} and the expression of the spin conductivity given in Eq.~\eqref{spinc}, we realize that the spin diffusion coefficient is fully determined if the single-particle spectral function is known. The single-particle spectral function of carriers can be calculated by solving self-consistently the set of Eqs.~\eqref{Green1} , \eqref{Green2}, and \eqref{Green3} of the DMFT. In the  present work, we focus the spin dynamical properties in the paramagnetic phase under a direct influence of magnetic impurities, i.e., in the condition $T\sim U\ll J$, where the temperature $T$ is chosen to be larger than a typical critical value of the paramagnetic-ferromagnetic transition temperature, $T_c=0.05$ (c.f. Ref.~\onlinecite{Chatt}). Hereafter, we mainly take $J=3$, $T=0.1$, and $U=0.5$, but a wide range of the local exchange coupling and temperature is also considered. Due to the heavy compensation in almost DMS systems, we use $n<x\ll 1$.~\cite{Das,HS05} The chemical potential therefore is located in the lower energy band edge, that is an important point characterizing the DMS systems.

\begin{figure}[t]
\includegraphics[angle=0,width=0.43\textwidth]{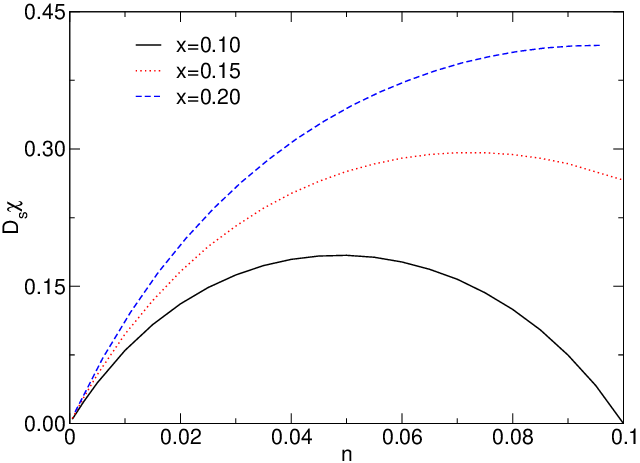}
\caption{(Color online) $D_s\chi$ vs carrier concentration $n$ at $J=3$, $U=0.5$, and $T=0.1$ for some densities $x$ of magnetic impurities.}
\label{f1}
\end{figure}

\begin{figure}[t]
\includegraphics[angle=-0,width=0.43\textwidth]{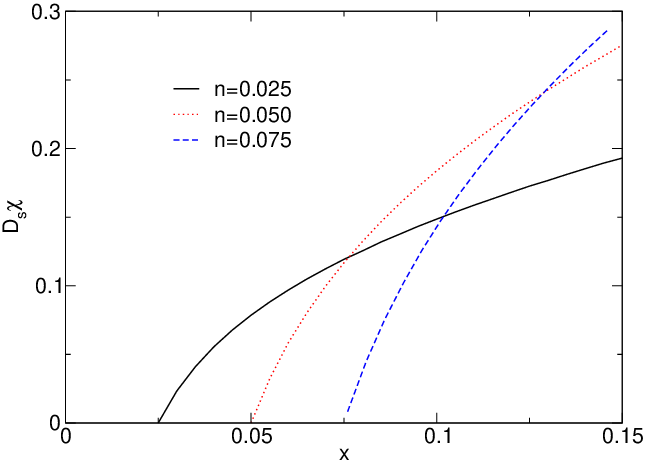}
\caption{(Color online) The dependence of $D_s\chi$ on $x$ at $J=3$, $U=0.5$, and $T=0.1$ for different values of $n$.}
\label{f3}
\end{figure}

\begin{figure}[t]
\includegraphics[angle=-0,width=0.43\textwidth]{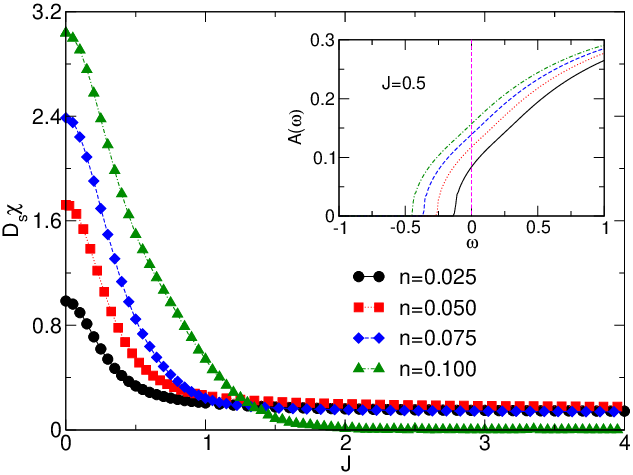}
\caption{(Color online) Dependence of $D_s\chi$ on $J$ at $U=0.5$, $x=0.1$, and $T=0.1$ for different values of $n$. The inset shows the DOS of itinerant carriers $A(\omega)$ at $J=0.5$ for the same different values of $n$ mentioned in the main panel, plotted in the same line styles. Here a vertical dashed-line denotes the Fermi level.}
\label{f2}
\end{figure}

Let us start with the spin diffusion characters in a variation of the carrier concentration $n$ and magnetic impurity density $x$. Figure~\ref{f1} displays the dependence of $D_s\chi$  on $n$ for several $x$ values at $J=3$, $T=0.1$, and $U=0.5$. Increasing $n$, $D_s\chi$ first increases and reaches its maximum at $n=x/2$, and then approaches zero once $n=x$ (see the solid line for $x=0.1$, for instance). In this case, the magnetic coupling is strong ($J=3$) so the impurity band is completely separated from the main band.~\cite{HS05} Our calculation (not shown here) reveals that a critical Hund coupling for a separation between the main and impurity bands is $J_c=1.2$. When varying the carrier occupation from zero to $x$, the chemical potential sweeps from $-\infty$ to the impurity level (acceptor energy level) which is isolated below the main band for $J>J_c$. The spectral weight becomes maximal at the center of the impurity band. Note here that the spin conductivity in a degenerate system normally depends linearly on the density of states at the Fermi level,~\cite{PR10} so when the Fermi level approaches the impurity band center, the spin conductivity would be enhanced and promotes the spin diffusion. The existence of spin diffusion indicates that the system is a normal spin conductor.~\cite{Kopi} When $n=x$, the impurity band is fully occupied, and in this case the chemical potential locates in the gap separated between the impurity band and the main band. Spin diffusion therefore is suppresses to zero and the system is a spin insulator. As mentioned above, this scenario only occurs for large $J$. In the opposite case, i.e.,~at small $J<J_c$, the gap that opened between the impurity band and the main band vanishes, the $D_s\chi$ at $n=x$ therefore is nonzero, because of nonzero concentration of the itinerant carriers. The behavior of spin diffusion in a wide range of the local exchange coupling can be seen in Fig.~\ref{f2}. When $x$ increases, the bandwidth of the impurity band increases since the number of states in the impurity band increases with $x$.~\cite{HS05} For a given value of the carrier concentration, the density of states at the Fermi level therefore is enhanced, leading to promote the spin conductivity or the spin diffusion in the system. Figure~\ref{f3} illustrates that property, and there we also present the spin diffusion as a function of $x$ for a given value of $n$. Clearly, one can see that the spin diffusion is completely zero if $x=n$ and monotonically increases as $x$ increases further.

In Fig.~\ref{f2}, we show the dependence of $D_s\chi$ on the local magnetic coupling $J$ at $x=0.1$ for different small concentrations $n$ at $T=0.1$ and $U=0.5$. For large magnetic couplings, we see that $D_s\chi$ is saturated and independent of $J$. This behavior looks similar to the result that emerged in the double exchange model in which Chernyshev and Fishman concluded that $D_s\chi$ is independent of $J$ if $J\gg 1$.~\cite{Cher} According to the above discussion of Fig.~\ref{f1}, $D_s\chi$ reaches zero if $n=x$ but is non-zero if $n<x$, as expected. In contrast, for small values of $J$, the inelastic magnetic scattering of itinerant carriers with local moments reduces, consequently it leads to an increase of the spin diffusion constant. In this regime, for a given value of $J$, $D_s\chi$ linearly increases with increasing itinerant carrier concentration. This is similar to the case of large $J$ discussed previously and presented in Fig.~\ref{f1} but only for $n<x/2$.
In order to illustrate the behavior of $D_s\chi$ in this case, we also show in the inset of Fig.~\ref{f2} the carrier density of states $A_\uparrow(\omega)=A_\downarrow(\omega)=A(\omega)$ at $J=0.5$ for the same values of $n$ in the main panel. Here we keep in mind that $D_s\chi$ is characterized by the density of states at the Fermi level. Inspecting the vertical line in the inset we can extract the magnitude of the carrier density of states at the Fermi level $A(E_F)$ for different $n$. Apparently, $A(E_F)$ increases if $n$ increases. This explains the behavior of the spin dynamics in the system for small $J$. Note here that the impurity band is not separated from the main band because the local magnetic coupling is small.~\cite{HS05} Thus in this case, the spin dependent potential is not sufficient to form a bound state. When $J$ is large enough, a hole with appropriate spin would be bound to a magnetic ion site. In the paramagnetic state, some hopping processes therefore are blocked,~\cite{Chat00} which diminish the spin diffusion. In the limit of $J\rightarrow \infty$, each carrier is bound to an Mn site and all hopping processes of the carriers in the systems are forbidden, i.e., in the case when the number of carriers is exactly equal to the number of magnetic impurities, the spin diffusion completely disappears. The spin diffusion is nonzero only if $n<x$.

\begin{figure}[t]
\includegraphics[angle=-0,width=0.43\textwidth]{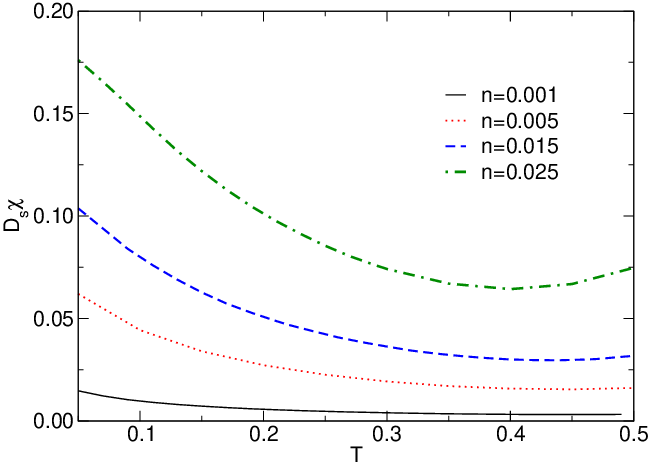}
\caption{(Color online) Dependence of $D_s\chi$ on temperature at $J=3$, $x=0.1$, and $U=0.5$ for different values of $n$.}
\label{f4}
\end{figure}

Finally, we present the temperature dependence of the spin dynamics in the paramagnetic DMS. In Fig.~\ref{f4}, we show the behavior of $D_s\chi$ versus temperature in the paramagnetic state at $J=3$, $U=0.5$, and $x=0.1$ for different carrier concentrations $n<x$. Increasing temperature always leads to enhancing the scattering between the carrier and impurities, thus decreasing the spin lifetime.~\cite{PR10} Consequently, as shown in Fig.~\ref{f4}, it lowers the spin diffusion constant. In contrast, lowering temperature promotes a quantum coherent of the magnetic order corresponding to favor the spin stiffness before the temperature reaches the Curie point. This behavior of $D_s\chi$ looks similar to that obtained in the $t-J$ model by using the high temperature Lanczos method.~\cite{Bon} Indeed, for small carrier concentrations ($n<0.005$), $D_s\chi$ displays a monotonic $1/T$-like temperature dependence. Increasing the carrier concentration, $D_s\chi$ increases, but for larger $n$, $D_s\chi (T)$ exhibits a minimum at an intermediate temperature.

The unusual nonmonotonic temperature dependence of $D_s\chi$ at large carrier densities can be explained if we attribute an existence of low-energy short lived many body states in the system.~\cite{Cher} The low-energy short lived many body states are formed as consequence of many-body effect in the paramagnetic state. Indeed, lowering temperature in the paramagnetic state might form spin density excitations existing at a short lived time. At low frequencies, these excitations become non-negligible if the carrier density is large enough.~\cite{JP95} According to the enhancement of the low-energy states, the inelastic magnetic scattering is enhanced and thereby the spin diffusion is suppressed. Close to the Curie temperature $T_c$, the static magnetic susceptibility $\chi$ rapidly increases, the spin diffusion constant therefore would be systematically decreased $D_s\sim 1/\chi$. Further information about the magnetic correlations within the inhomogeneous paramagnetic state in DMS systems therefore might be observed by analyzing the measurements of the spin diffusion constant and the static magnetic susceptibility.~\cite{Dai01}

\section{Conclusion}
We have discussed the spin dynamic scenario in paramagnetic diluted magnetic semiconductors within the dynamical mean-field theory which is exactly solvable in the infinite dimensional limit. The single particle Green function of the Kondo lattice model including the random disorder thereby has been explicitly calculated. By employing the exact spectral representation for the spin-spin correlation function and in the hydrodynamic regime that can be applied to the spin relaxation in the presented model, we have derived the general Einstein relation between the spin diffusion coefficient and the spin conductivity. Following the Greenwood formalism, the spin conductivity has been expressed via the single-particle spectral function, which allows us to calculate the spin diffusion constant based on the dynamical mean field theory. It is found that the spin diffusion enhances if the Fermi level settles inside the magnetic impurity band. Both magnetic scattering and temperature suppress the spin diffusion in the paramagnetic state. The existence of the minimum point in the $D_s\chi(T)$ curve for high enough carrier dopings has been explained by the occurrence of low-energy short lived many body states in the system. The influence of random disorder in the spin diffusion in the DMSs is also important, however we will leave its consideration to the future.

\acknowledgements
This research is funded by the Vietnam National Foundation for Science and Technology Development (NAFOSTED) under grant No 103.01-2014.05.

\end{document}